\PassOptionsToPackage{dvipsnames}{xcolor}
\documentclass[sigplan,screen]{acmart}

\usepackage{graphicx}
\usepackage{booktabs}
\usepackage{array}
\usepackage{amsfonts}
\usepackage{nicefrac}
\usepackage{microtype}
\usepackage{subcaption}
\usepackage{caption}
\usepackage{adjustbox}
\usepackage{xspace}
\usepackage{enumitem}
\usepackage{listings}
\usepackage{amsmath}
\usepackage[T1]{fontenc}
\usepackage{multirow}
\usepackage{fvextra}
\usepackage{soul}
\usepackage{diagbox}
\usepackage{ragged2e}
\usepackage{fancyvrb}
\usepackage{float}
\usepackage{balance}
\usepackage{multicol}

\newcommand{\mypara}[1]{
\vspace{6pt}
\noindent
\textbf{#1}.
}

\newcommand{\mylist}[1]{
\vspace{4pt}
\noindent
\textbf{#1}.
}

\newcommand{\repobench}{\texttt{RVBench}\xspace}
\newcommand{\verusbench}{\texttt{VerusBench}\xspace}
\newcommand{\ragverus}{\textsc{RagVerus}\xspace}
\newcommand{\autoverus}{\textsc{AutoVerus}\xspace}
\newcommand{\verismo}{\texttt{VeriSMo}\xspace}

\newcommand{\ironkv}{\texttt{IronKV}\xspace}

\definecolor{codegreen}{rgb}{0,0.6,0}
\definecolor{codepurple}{rgb}{0.58,0,0.82}
\definecolor{codeyellow}{rgb}{1,0.96,0.7}
\definecolor{codegray}{rgb}{0.5,0.5,0.5}
\definecolor{lightcoral}{rgb}{1,0.9,0.8}
\definecolor{backcolour}{rgb}{0.95,0.95,0.92}

\setcopyright{cc}
\setcctype{by}
\acmDOI{10.1145/3759425.3763382}
\acmYear{2025}
\copyrightyear{2025}
\acmISBN{979-8-4007-2148-9/25/10}
\acmConference[LMPL '25]{Proceedings of the 1st ACM SIGPLAN International Workshop on Language Models and Programming Languages}{October 12--18, 2025}{Singapore, Singapore}
\acmBooktitle{Proceedings of the 1st ACM SIGPLAN International Workshop on Language Models and Programming Languages (LMPL '25), October 12--18, 2025, Singapore, Singapore}
\received{2025-07-05}
\received[accepted]{2025-08-08}

\date{July 2025}

\begin{document}

\title{Towards Repository-Level Program Verification with Large Language Models}

\author{Si Cheng Zhong}
\orcid{0009-0000-0697-0532}
\affiliation{%
    \institution{University of Toronto}
    \city{Toronto}
    \state{Ontario}
    \country{Canada}
}
\email{sicheng.zhong@mail.utoronto.ca}

\author{Xujie Si}
\orcid{0000-0002-3739-2269}
\affiliation{%
    \institution{University of Toronto}
    \city{Toronto}
    \state{Ontario}
    \country{Canada}
}
\email{six@cs.toronto.edu}




\begin{abstract}

Recent advancements in large language models (LLMs) suggest great promises in code and proof generations.
However, scaling automated formal verification to real-world projects requires resolving cross-module dependencies and global contexts, which are crucial challenges overlooked by existing LLM-based methods with a special focus on targeting isolated, function-level verification tasks.
To systematically explore and address the significant challenges of verifying entire software repositories, we introduce \repobench, the first verification benchmark explicitly designed for repository-level evaluation, constructed from four diverse and complex open-source Verus projects. 

We further introduce \ragverus, an extensible framework that synergizes retrieval-augmented generation with context-aware prompting to automate proof synthesis for multi-module repositories.
\ragverus \textit{triples} proof pass rates on existing benchmarks under constrained model inference budgets, and achieves a \textit{27\%} relative improvement on the more challenging \repobench benchmark, demonstrating a scalable and sample-efficient verification solution.\footnotetext{The benchmark and experimental artifact are publicly available at \url{https://github.com/GouQi12138/RVBench}}



\end{abstract}




\maketitle

\section{Introduction}

High-assurance software, such as operating systems and financial infrastructure, demands rigorous correctness guarantees~\cite{li2024survey,zhang2024selene}. Formal verification replaces testing with mathematical proofs that a program adheres to its specifications, offering exhaustive guarantees over all possible executions. 
Languages designed with verification in mind, such as Verus~\cite{lattuada2023verus} and Dafny~\cite{leino2010dafny}, leverage integrated language features and automated constraint solving to facilitate this process. Nevertheless, constructing proofs remains labor-intensive, requiring extensive human expertise to address complex invariants, lemma interactions, and solver constraints~\cite{wen2024enchanting,lattuada2024verusExp,li2023veriScale}. This expertise bottleneck intensifies in large-scale, real-world repositories, where proofs span across multiple interdependent modules, significantly complicating proof construction and premise selection.

Recent advancements in large language models (LLMs) have shown promising potential to alleviate the verification workload by automating proof synthesis~\cite{kamath2023loopy,sun2024clover}. Methods leveraging LLMs iteratively generate candidate proofs, refine them based on compiler feedback, and adapt their training processes accordingly~\cite{yang2024autoverus,loughridge2024dafnyBench,chen2024evolveVerus,aggarwal2024alphaverus,chakraborty2024FStar}. Despite these advancements, current research predominantly targets isolated, function-level tasks with limited complexity, overlooking the substantial challenges inherent in repository-level verification.

Repository-level program verification introduces unique difficulties due to extensive codebase sizes, intricate cross-module dependencies, and custom-defined language constructs specific to each project. These complexities require the verifier to effectively navigate a large context, identify relevant lemmas, and adhere strictly to project-specific conventions—tasks that challenge both human verifiers and automated LLM-based systems.

To systematically evaluate and address these challenges, we introduce \repobench, the first benchmark specifically designed for repository-level verification tasks. \repobench is curated from four diverse and award-winning open-source Verus projects~\cite{zhou2024verismo,sun2024anvil,hawblitzel2015ironkv,cai2024vest}, reflecting real-world verification challenges, including complex proof structures, inter-module lemma dependencies, and custom type definitions. This benchmark consists of 755 proof completion tasks across 337 modules and 3,464 functions, significantly scaling the complexity compared to existing function-level benchmarks (i.e., 150 functions used in \verusbench~\cite{yang2024autoverus}).

We further propose \ragverus, a simple retrieval-based framework tailored for repository-scale verification, which dynamically integrates contextually relevant examples and dependencies from extensive codebases into the LLM's reasoning process. Although retrieval-augmented generation (RAG) methods are well-established in natural language processing (NLP) and machine learning literature~\cite{karpukhin2020dense,NEURIPS2020_6b493230}, \ragverus's distinct contribution lies in applying and evaluating these methods explicitly for the nuanced demands of repository-level verification tasks.

Experimental evaluations demonstrate that \ragverus significantly outperforms prior frameworks like \autoverus~\cite{yang2024autoverus} on both function-level and, more importantly, repository-level tasks. Our results highlight the critical role of hybrid retrieval strategies in navigating repository complexities, marking a substantial step towards scalable, automated verification methods.

In summary, this paper makes the following contributions:
\begin{itemize}[leftmargin=15pt]
\item We introduce \repobench, the first benchmark for systematically evaluating repository-level verification in Verus, addressing a crucial gap in existing verification research.
\item We propose \ragverus, a retrieval-based verification framework that leverages structured repository metadata to enhance LLM-driven proof synthesis.
\item We conduct thorough experimental evaluations illustrating the advantages and current limitations of retrieval-based approaches to repository-level verification, thus setting a strong baseline for future improvements.
\end{itemize}

\section{Background}

\subsection{The Verus Verifier}

Verus~\cite{lattuada2023verus} is a \textit{verification-aware language} and SMT-based verification tool designed as an extension to Rust, an imperative programming language. It integrates Rust's strong ownership and type systems with formal methods, enabling developers to enforce logical correctness alongside memory safety. Verus introduces a domain-specific language (DSL) for embedding \textit{specification}, \textit{proof annotations}, and \textit{executable code} inline, clearly delineating these three distinct code modes (an illustrative example is shown in Figure~\ref{fig:verus_code}):

\begin{enumerate}
    \item \texttt{spec} mode: Defines preconditions, postconditions, and logical requirements \\ e.g., {\tt \textbf{requires} x > 0}, \, {\tt \textbf{ensures} result > 0}.
    \item \texttt{proof} mode: Provides proof annotations, assertions, loop invariants, and lemma proofs necessary for bridging specifications to implementations \\e.g., {\tt assert\_by(x > 0, \{..\})},\, {\tt \textbf{invariants} i < N}.
    \item \texttt{exec} mode: Contains standard executable Rust code, compiled and executed normally, but checked against specifications at compile time.
\end{enumerate}

The specifications and proofs defined in Verus are ghost annotations, erased at runtime but essential for static verification. This structured separation ensures clarity between functional logic and runtime behavior, aiding automated verification.

\begin{figure}[h!]
\begin{lstlisting}[frame=none]
// Returns the greatest_lower_bound as evidence for the proof of correctness for the set data structure
fn get(&self, k: &K) -> (res: (ID, Ghost<KeyIterator<K>>))
\end{lstlisting}
\vspace{-0.7\baselineskip}
\begin{lstlisting}[frame=none, firstnumber=3, backgroundcolor=\color{codeyellow}]
   requires
      self.valid(),
   ensures ({
      let (id, glb) = res;
      &&& id@ == self@[*k]
      &&& self.lows.greatest_lower_bound_spec(KeyIterator::new_spec(*k), glb@)
      &&& id@.valid_physical_address()
   }),
\end{lstlisting}
\vspace{-0.7\baselineskip}
\begin{lstlisting}[frame=none, firstnumber=11]
{
   let ki = KeyIterator::new(k.clone());
   let glb = self.lows.greatest_lower_bound(&ki);
\end{lstlisting}
\vspace{-0.7\baselineskip}
\begin{lstlisting}[frame=none, firstnumber=14, backgroundcolor=\color{lightcoral}]
   proof {
      let glb_k = *glb.get();
      assert(self.lows@.contains_key(glb_k)); // OBSERVE
      let hi = choose |hi| self.lows.gap(glb, hi) && #[trigger] KeyIterator::between(glb, ki, hi);
      assert(KeyIterator::between(KeyIterator::new_spec(glb_k), ki, hi));
      assert(self.lows@.contains_key(glb_k)
          && self.lows.gap(KeyIterator::new_spec(glb_k),hi)
          && KeyIterator::between(KeyIterator::new_spec(glb_k), KeyIterator::new_spec(*k), hi));
   }
\end{lstlisting}
\vspace{-0.7\baselineskip}
\begin{lstlisting}[frame=none, firstnumber=23]
    let id = (*self.lows.get(glb.get()).unwrap()).clone_up_to_view();
    (id, Ghost(glb))
}
\end{lstlisting}
    \caption{An annotated Verus function with \texttt{spec} in \sethlcolor{codeyellow}\hl{yellow} and \texttt{proof} in \sethlcolor{lightcoral}\hl{orange}. Note that each code mode can have \textit{function calls} requiring dependency resolution from other sources.}
    \label{fig:verus_code}
\end{figure}

Verus translates annotated Rust code into satisfiability modulo theories (SMT) formulas, subsequently verified by SMT solvers like Z3~\cite{de2008z3}. Proof annotations guide the solver through complex scenarios~\cite{cho2024framework}, helping resolve ambiguities and bridging automation gaps during program verification.

\subsection{Large Language Models (LLMs) in Verification}

Recent advancements in generative AI, particularly large language models (LLMs) such as GPT-4~\cite{openai_gpt4o} and Gemini~\cite{gemini}, offer promising avenues for automating formal verification processes. These transformer-based architectures, trained on vast corpora of code, documentation, and mathematical texts, possess sophisticated probabilistic representations of syntax, semantics, and logical reasoning patterns.

Effective use of LLMs in verification typically involves strategies such as \textit{chain-of-thought (CoT)} prompting and \textit{retrieval-augmented generation (RAG)}. CoT prompting guides models to systematically decompose proofs into explicit, logical sub-steps (e.g., "first, assert loop termination; next, apply Lemma identity..."), mimicking human expert reasoning. However, while CoT hints useful reasoning patterns, it does not address complex cross-module dependencies that are common in repository-level tasks.

To address this limitation, RAG methods dynamically retrieve contextually relevant knowledge, such as code snippets, lemma definitions, and project-specific conventions, to enrich the LLM's reasoning context. While early retrieval-based approaches primarily utilized syntactic matching, recent innovations incorporate semantic encoders that capture nuanced code relationships and structural dependencies~\cite{yang2023leandojo, shrivastava2023repoPrompt}. By embedding repository-specific metadata and inter-module contexts, these methods significantly enhance LLM accuracy and efficiency in generating verifiable proofs across extensive codebases.

\section{\repobench: Creating A New Repository-level Verification Benchmark}

Although existing benchmarks such as \verusbench have been instrumental in evaluating basic proof synthesis for isolated functions, their focus on self-contained tasks involving simple arithmetic reasoning and relevant invariants has led to near saturation~\cite{yang2024autoverus}. These benchmarks lack the complexities inherent in real-world software, particularly cross-module dependencies and environmental interactions. To address this gap, we introduce \repobench, the first repository-level benchmark for verification-aware languages, designed to evaluate tools on full-scale codebases mirroring industrial practices.

\subsection{Benchmark Construction and Scope}

To ensure diverse, real-world challenges reflective of practical verification tasks, \repobench is curated from four prominent open-source Verus projects, each selected for its complexity, domain-specific logic, and inter-module dependencies:

\begin{enumerate}
    \item \textbf{VeriSMo}~\cite{zhou2024verismo}: A verified security module for confidential virtual machines (VMs), ensuring memory access safety and confidentiality via permission-based reasoning.
    \item \textbf{Anvil}~\cite{sun2024anvil}: A framework for verifying Kubernetes controller correctness and liveness, leveraging Verus to enforce state-machine invariants in cloud-native systems.
    \item \textbf{IronKV}~\cite{hawblitzel2015ironkv}: A distributed key-value store with proofs for consensus protocols and network fault tolerance.
    \item \textbf{Vest}~\cite{cai2024vest}: A formally verified set of high-performance parsers and serializers for binary data formats.
\end{enumerate}

Together, these repositories provide a rich set of 3,464 functions spread across 337 modules, forming a robust testbed for evaluating tools at the repository scale. We specifically derive 755 proof completion tasks emphasizing cross-module reasoning and project-specific dependencies.


\subsection{Benchmark Creation Methodology}

We established a rigorous methodology to ensure \repobench accurately captures the complexity of repository-level verification tasks. Specifically, we process Verus repositories into structured datasets through a two-stage pipeline: identifying taskable locations in individual functions, followed by static syntax analysis to build supporting metadata contexts for all functions and constructs.


\subsubsection{Task Identification}
We build a new line-parse tool leveraging Verus' compiler and  code-mode parser to identify formal properties such as function modes, specification conditions, and proof invariants. By precisely controlling over the \textbf{code mode} hierarchy, we enable flexible creation of diverse evaluation tasks. For instance, proof completion tasks are generated by masking proof annotations (e.g., loop invariants, assertions) from verified functions and modules to form task queries.

\subsubsection{Metadata Extraction}
We extract repository-wide metadata from Rust codebases to index individual functions under the contextualized verification task, informed by context modeling in code generation~\cite{shrivastava2023repoPrompt}. The metadata indexing each function contains: i) the file name, any applicable construct name, and the function name; ii) the function's type signature; iii) method invocations; iv) type identifiers and variable declarations; and v) the function's code mode. Through static analysis over the current file, related parent/child structures, and imported modules, we model a linkage graph that captures control and data flow, revealing necessary premise relationships such as which \texttt{proof} function invokes which \texttt{spec} lemmas. This structured indexing enables precise evaluation of context-aware verification, where an agent must resolve interconnected dependencies and project-specific semantics in an uninformed situation.

\subsection{Proof Completion Tasks}

We identify functions containing proof annotations from the parsed repositories and form task queries as:
\begin{itemize}
    \item \textbf{Input}: A \texttt{proof} function or \texttt{exec} function that has complete specification lines (pre/post-conditions) but with all \texttt{proof} lines omitted.
    \item \textbf{Expected output}: A function together with proof annotations to \textit{replace} the input, ideally ensuring syntactic and semantic correctness of the problem function.
\end{itemize}
The generated answer is inserted back into the codebase to compile for verification.
We exclude all tasks where Verus can successfully verify even without proof annotations, such as trivial arithmetic properties resolvable by Z3's default tactics.

\begin{table*}[ht]
\centering
\caption{\repobench Composition across Four Verus Projects and Comparison with \verusbench}
\setlength{\tabcolsep}{8pt} 
\renewcommand{\arraystretch}{0.8} 
\adjustbox{max width=2\textwidth}{
\begin{tabular}{lrrrrr}
\toprule
\textbf{Source} & \textbf{Type} & \multirow{2}{*}{\textbf{Functions}} & \textbf{Proof} & \textbf{Proof} & \textbf{Complex} \\
& \textbf{Constructs} && \textbf{Tasks} & \textbf{Lines} & \textbf{Tasks} \\

\midrule
\textbf{\verusbench}\cite{yang2024autoverus} & - & (150) & (150) & ($\sim$1,200) & (0) \\

\midrule
\midrule
VeriSMo & 224 & 1,656 & 383 & 7,962 & 331 \\
\midrule
Vest: & 45 & 1,013 & 214 & 1,227 & 198 \\
\ \ \ \ Vest-core & 35 & 739 & 179 & 1,055 & 164 \\
\ \ \ \ Vest-example & 10 & 274 & 35 & 172 & 34 \\
\midrule
IronKV & 34 & 528 & 129 & 2,377 & 117 \\
\midrule
Anvil (executable): & 34 & 267 & 29 & 479 & 27 \\
\ \ \ \ Anvil-fluent & 9 & 64 & 7 & 106 & 6 \\
\ \ \ \ Anvil-rabbitmq & 14 & 112 & 9 & 175 & 9 \\
\ \ \ \ Anvil-replicaset & 2 & 17 & 5 & 106 & 4 \\
\ \ \ \ Anvil-zookeeper & 9 & 74 & 8 & 92 & 8 \\
\midrule
\textbf{\repobench Total} & 337 & 3,464 & 755 & 12,045 & 673 \\
\bottomrule
\end{tabular}
}
\label{tab:benchmark}
\end{table*}

Table~\ref{tab:benchmark} summarizes important statistics of \verusbench (second row), verification tasks used in the previous work~\cite{yang2024autoverus}, and the newly constructed \repobench (third to seventh row), which consists of four Verus projects. 
Taking the \verismo project as an example, we extracted \textbf{224} project-specific type constructs and 2,073 code pieces, including macro definitions and functions, and parsed \textbf{1,656} indexable functions, out of which we identified 460 functions containing \texttt{proof} lines in the original code using the syntax tracer. We further filtered out functions that the Verus compiler can automatically solve even without proof code, which resulted in \textbf{383} tasks for the benchmark, consisting of \textbf{7,962} lines of proof in total.


Furthermore, we carefully analyze the ground-truth proofs and split the benchmark into two categories based on whether the proofs require premises (aka proof function dependencies):
\begin{itemize}
    \item \textbf{Simple tasks}: Isolated functions whose proofs require project-specific macros and type definitions but no function calls. All tasks of \verusbench fall in this category.
    \item \textbf{Complex tasks}: Functions whose proofs require complicated premises, i.e., dependencies of other proof functions.
\end{itemize}
\noindent
\textbf{Simple tasks} test \textit{contextual understanding} of project-specific syntax (e.g., custom defined macros, types and constructs). For instance, a function may use a custom integer type (\texttt{SecureInt}) defined in a project header; understanding this type is essential for proof syntax, even if no premises are invovled. \textbf{Complex tasks} evaluate \textit{premise retrieval} and \textit{integration} (i.e., resolving inter-module proof functions). This categorization isolates distinct repository-level challenges, where \repobench differentiates localized syntax adaptation and systemic dependency resolution to provide a granular framework for diagnosing agent limitations.


\begin{table}[htbp]
    \centering
    \caption{Logic and Problem Characteristics of Proof-Completion Tasks: \textbf{LL}-Linear Logic, \textbf{SL}-Separation Logic, \textbf{TL}-Temporal Logic, \textbf{Cc}.-Concurrency, \textbf{Qt}.-Quantifiers}
    \begin{tabular}{|l| *{5}{c|}}
        \toprule
        \textbf{Source} & \textbf{LL} & \textbf{SL} & \textbf{TL} & \textbf{Cc.} & \textbf{Qt.} \\
        \midrule
        \multicolumn{6}{|l|}{\textbf{\repobench Sources}} \\
        \hline
        VeriSMo & $\checkmark$ & $\checkmark$ & $\times$ & $\checkmark$ & $\checkmark$ \\
        \hline
        IronKV & $\times$ & $\times$ & $\checkmark$ & $\checkmark$ & $\checkmark$ \\
        \hline
        Anvil & $\times$ & $\times$ & $\checkmark$ & $\checkmark$ & $\checkmark$ \\
        \hline
        Vest & $\checkmark$ & $\times$ & $\times$ & $\times$ & $\checkmark$ \\
        \hline
        \multicolumn{6}{|l|}{\textbf{Previous Benchmark}} \\
        \hline
        VerusBench & $\times$ & $\times$ & $\times$ & $\times$ & $\checkmark$ \\
        \bottomrule
    \end{tabular}
    \label{tab:logic_comparison}
\end{table}

\repobench also introduces a significant increase in the logical reasoning complexity required for proof writing as highlighted in Table \ref{tab:logic_comparison}.
For instance, verifying memory access confidentiality in VeriSMo necessitates reasoning with linear logic to track resource consumption and separation logic to handle disjoint memory. IronKV and Anvil's distributed consensus and fault tolerance aspects require temporal and concurrency reasoning, utilizing quantifiers~\cite{sun2024anvil}. Unlike \verusbench's isolated context, \repobench demands reasoning about modularity, resources, temporal evolution, and concurrency, reflecting real-world verification challenges.

\subsection{Evaluation Metrics}

\repobench employs the following metrics and infrastructure to evaluate proof completion quality:

\begin{figure*}
    \centering
    \includegraphics[width=0.8\textwidth]{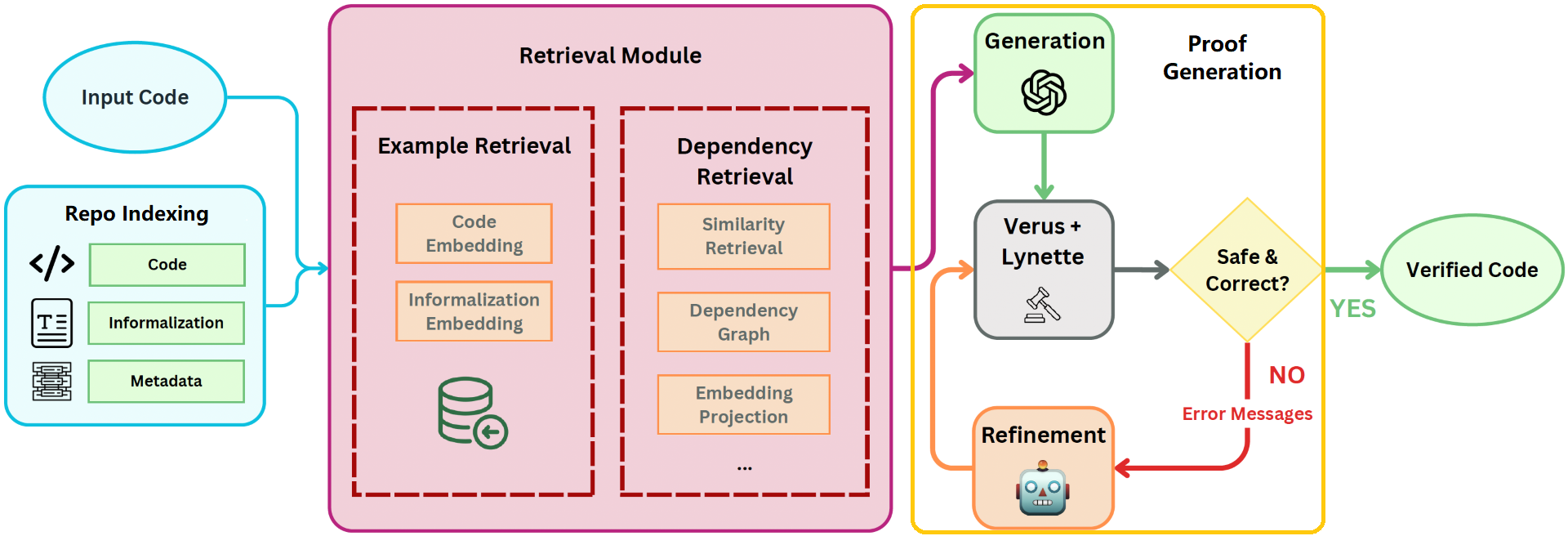}
    \caption{The \ragverus pipeline consists of three stages: 1) mining code properties, 2) retrieving task-specific information, and 3) generating proofs for verifiable code}
    \label{fig:RAG_Verus}
\end{figure*}

\begin{itemize}[leftmargin=20pt]
    \item \textbf{Correct}: The code is correct if the generated proof annotations pass Verus compilation, which verifies logical soundness across the entire project, validating the required specifications and any propagated constraints.
    \item \textbf{Intact}: The generated result is intact if the executable code and specifications remain unaltered, preserving the original functionality for verification. We check this by the \textbf{\textit{Lynette}} syntax checker~\cite{yang2024autoverus}, which validates code consistency for all \texttt{exec} and \texttt{spec} lines (e.g., no injected \texttt{ensures} or modified variable types). This metric is important because LLMs could simply overwrite the executable code and/or specifications, making generated proofs irrelevant regardless being correct or not.
    \item \textbf{Success}: The aggregated success metric requires a result to be both correct and intact. For example, a memory allocator proof that compiles (correct) but inadvertently modifies a pointer type (contaminated) is deemed a failure.
\end{itemize}

As verification success is a rigid binary metric, we also introduce a \textit{relaxed} metric, which compares the generated code and proofs to the ground-truth reference using the \textbf{BLEU score} (Bilingual Evaluation Understudy) widely used in NLP tasks, measuring textual similarity between the two, to assess alignment with expected coding standards and term usages, thus signaling gradual improvement in proof generation.


\section{\ragverus Framework}


To effectively address the complexities inherent in repository-level verification, we introduce \ragverus, a retrieval-augmen-ted generation (RAG) framework explicitly designed to leverage large language models (LLMs) for program verification at scale. 
As illustrated in Figure~\ref{fig:RAG_Verus}, \ragverus consists of three modules --- repository indexing module, retrieval module, and proof generation module. 
Given that the proof generation module largely follows the design of the previous work~\cite{yang2024autoverus}, we will primarily illustrate the first two modules. 


\subsection{Repository Indexing}

We prepare persistent indices for efficient retrieval of all contents within a verification task repository. This involves preprocessing: 1) code for all functions, and 2) natural language summaries derived from function metadata.

\mypara{Code Indexing}
For efficient retrieval of relevant verification contexts, we use persistent vector stores where code documents are encoded into high-dimensional embeddings using OpenAI's \texttt{text-embedding-3-large} model~\cite{openai2024embedding}, and are then retrieved via FAISS~\cite{faisspaper}, a fast indexing library. This approach allows us to capture the syntax and structures of code snippets, going beyond simple keyword matching.

\mypara{Metadata Indexing}
While statically extracted formal metadata provides a structured view of the existing codebase, it is unable to recover the missing premises of an unverified task function. Moreover, the high-level intent or semantic relationships between code elements are not always explicitly captured by static analysis.

To find the most relevant dependencies, we further \textit{informalize} code and associated metadata into natural language descriptions (see Appendix~\ref{appendix:data} for more details)  and then index natural language descriptions. This informal natural language summary captures code behavior beyond syntax and allows us to retrieve functionally dependent examples, ensuring the verification process is guided by relevant context that aligns with the task's high-level intent.

\vspace{4pt}
To prevent potential data leakage, both indexing systems exclude the target function’s own verified context (if present) during retrieval, ensuring proofs are synthesized solely from external dependencies for the unverified function.


\subsection{Retrieving Useful Contexts for Repository-level Verification}

\ragverus employs two complementary retrieval strategies to enhance proof synthesis:
1) few-shot example retrieval, and 2) dependency retrieval. Retrieved contexts are integrated into the prompts to guide LLMs in generating verifiable code, balancing structural patterns (via examples) and logical prerequisites (via dependencies). The framework’s modular design supports customized retrieval models to tailor specific verification challenges.

\subsubsection{Few-Shot Example Retrieval}

To guide LLMs to adhere to correct styles and practices, \ragverus uses few-shot retrieval to select representative \textit{problem-approach} pairs from other verified examples. Using the tasked \textit{code} or its informalized \textit{metadata} as query, we retrieve syntactically or semantically similar, proof-masked functions paired with their verified counterparts, forming \textit{reference solution pairs}.

To cover diverse proof styles, we source from multiple indices, including tutorial examples from the official Verus repository, tasks from \verusbench and \repobench, among many others. After ranking the retrieved candidates, we retain a maximum of three diverse and relevant examples per task query. These examples provide the LLM with contextual references for the proof-completion task~\cite{brown2020fewshot}, improving verification quality and precision.

\subsubsection{Dependency Retrieval}

Dependency retrieval in \ragverus identifies cross-module function signatures, prem-ises, and type definitions necessary to inform proof synthesis in large codebases while adhering to LLM context windows. This process addresses two critical needs:
\begin{enumerate}[leftmargin=20pt]
    \item \textbf{Context Awareness}: Integrates project-specific constructs, type definitions, and code conventions to ground proofs in repository-wide patterns and knowledge.
    \item \textbf{Premise Retrieval}: Identifies premises, i.e., functions in \texttt{spec} or \texttt{proof} modes, that are potentially useful during the verification of the target task.
\end{enumerate}

Well-chosen dependency hints resolve logical gaps in verification conditions while filtering noise, providing contexts on module interactions within the repository. Identifying a \textit{formatted list of verified dependent functions and constructs} creates an essential toolkit for the LLM to elaborate on, streamlining the generation process and encouraging function reuse.

We evaluate a simple embedding-based approach using informalized metadata similarity to capture context and dependency relationships with the tasked function, covering both the related definitions and the dependent premises.




\subsection{Proof Generation}

In \ragverus, proof generation involves synthesizing verification annotations such as loop invariants, quantifier logic, and termination conditions to link \texttt{spec} and \texttt{exec} modes explicitly. Given a partially annotated task function, the LLM generates complete, executable Verus annotations informed by the retrieved contexts. Generated proofs are rigorously validated using:
\begin{itemize}
    \item \textbf{Verus Compiler}: Ensures logical correctness and compliance with the repository-wide specifications.
    \item \textbf{Lynette Syntax Checker}: Validates intactness, ensuring no unintended modifications occur in the executable or specification code.
\end{itemize}

If a proof fails initial validation, \ragverus optionally supports iterative refinement cycles, leveraging feedback from the verifier to iteratively enhance proof correctness.

\section{Evaluation}

We conducted comprehensive evaluations of \ragverus to rigorously assess its effectiveness in handling both function-level and repository-level verification tasks. These evaluations demonstrate the practical benefits as well as highlight the current limitations of retrieval-augmented verification.

\subsection{Baseline Methods \& Setup}

\mypara{Direct Proof Generation}
By leveraging LLMs' flexibility in code manipulation and by prompting the model with static Verus knowledge and expert-tuned instructions, we sample independent proof candidates that adhere to Verus' strict verification syntax. Given a single unverified Verus function, the LLM rewrites the entire file to insert complete proofs directly within the code using Verus' DSL, encompassing invariants, assertions, and standard lemma calls.

\mypara{Ensemble Refinement}
We improve the generated proof candidates by iteratively refining and repairing the verification \hspace{-0.1pt}annotations \hspace{-0.1pt}base \hspace{-0.1pt}on \hspace{-0.1pt}three \hspace{-0.1pt}approaches \hspace{-0.1pt}in \hspace{-0.1pt}\autoverus~\hspace{-1pt}\cite{yang2024autoverus}:
\begin{itemize}
    \item Candidate Merging: Identifies and merges sound proofs from multiple candidate variants.
    \item Refine Templates: Applies rule-based corrections (e.g., fixing type mismatches in ghost code, changing regular \texttt{u64} to ghost \texttt{integer}).
    \item Adaptive Repair: Instructs LLM repairs with elicited guiding examples for each kind of failure case (e.g., lacking invariants, syntax mismatches).
\end{itemize}

\noindent\autoverus using refinement, achieved near-perfect automation \hspace{-0.2pt}on \hspace{-0.2pt}function-level \hspace{-0.2pt}tasks - 91\% \hspace{-0.2pt}pass \hspace{-0.2pt}rate \hspace{-0.2pt}on \hspace{-0.2pt}\verusbench~\hspace{-1.3pt}\cite{yang2024autoverus}.

\subsection{Retrieval-Augmented Methods}

We augment upon the two baselines by incorporating two retrieval strategies: 1) few-shot retrieval and 2) dependency retrieval, to evaluate the effectiveness of \ragverus.

To ensure a fair comparison, we maintain an equal length of prompt instructions across all methods for the same task. The baselines are prompted with generic static examples, and RAG methods have dynamic examples. RAG methods only gain a potential advantage in context length when dependency retrieval is performed, up to 1000 more tokens.

We chose GPT-4o (version 2024-08-06) \cite{openai_gpt4o}, the most well-rounded model offered by OpenAI at the time of experiment as our model of choice for all baselines and augmented methods. The temperature was set to 1.0 for all sampling processes.

\subsection{Evaluation on Function-Level Verification Tasks}


\verusbench comprises 150 function-level proof-completion tasks, demanding arithmetic reasoning and relevant invariants.
We tested 139 available tasks from three sources\footnote{\autoverus was also tested on one unpublished Verus-CloverBench~\cite{sun2024clover}}: Diffy\\~\cite{chakraborty2021diffy}, MBPP~\cite{misu2024MbppDfy}, and Verus tutorials.

Since \verusbench does not have dependencies to consider, we evaluated the isolated impact of adaptive few-shot retrieval.
As the full refinement pipeline saturates this benchmark, we evaluated based on the direct generation pipeline, with and without retrieval augmentation. Our retrieval pool consisted of indexed files from \verusbench itself, official Verus tutorials, and example files used in \autoverus' refinement phase. We limited sampling to five attempts per task to assess the benefits of different retrieval strategies within a constrained budget. Table \ref{tab:Verus_bench_result_percent} presents the performance comparison, where RAG-Code and RAG-Text denote retrieval based on code similarity and informalized-summary similarity, respectively.


\begin{table}[ht!]
\centering
\caption{Evaluation results on \verusbench with percentages based on total number of tasks in each category.}
\setlength{\tabcolsep}{1pt}
\label{tab:Verus_bench_result_percent}
\begin{tabular}{@{}ll@{}ccc@{}}
\toprule
\textbf{Task}  & \textbf{Method}   & \textbf{Correct(n,\%)} & \textbf{Intact(n,\%)} & \textbf{Success(n,\%)} \\
\midrule
 & DirectGen        & 23 (29.5\%)         & 60 (76.9\%)      & 17 (21.8\%)         \\
\textbf{MBPP} & RAG-Code       & 57 (73.1\%)         & 74 (94.9\%)      & 54 (69.2\%)         \\
& RAG-Text    & 49 (62.8\%)         & 68 (87.2\%)      & 45 (57.7\%)         \\
\midrule
& DirectGen        & 3 (7.9\%)           & 30 (78.9\%)      & 2 (5.3\%)           \\
\textbf{Diffy} & RAG-Code       & 19 (50.0\%)         & 38 (100.0\%)     & 19 (50.0\%)         \\
& RAG-Text    & 24 (63.2\%)         & 37 (97.4\%)      & 23 (60.5\%)         \\
\midrule
& DirectGen        & 7 (30.4\%)          & 21 (91.3\%)      & 6 (26.1\%)          \\
\hspace{-1pt}\textbf{Tutorial} & RAG-Code       & 11 (47.8\%)         & 22 (95.7\%)      & 11 (47.8\%)         \\
& RAG-Text    & 8 (34.8\%)          & 22 (95.7\%)      & 8 (34.8\%)          \\
\midrule[0.8pt]
& DirectGen        & 33 (23.7\%)         & 111 (79.9\%)     & 25 (18.0\%)         \\
\textbf{Total} & RAG-Code       & \textbf{87 (62.6\%)} & 134 (96.4\%)     & \textbf{84 (60.4\%)} \\
& RAG-Text    & \textbf{81 (58.3\%)} & 127 (91.4\%)     & \textbf{76 (54.7\%)} \\
\bottomrule
\end{tabular}
\end{table}

RAG-Code consistently achieved higher performance across all metrics, particularly excelling on MBPP. Retrieval using informalized summaries also significantly improved upon the baseline, notably outperforming code-based retrieval in Diffy (23 vs. 19). Notably, our RAG results with limited sampling (5 attempts) already outperformed the same baseline reported in the \autoverus paper (44.7\% success rate using up to 125 invocations), highlighting the value of task-specific context over fixed guidance.

While code-based retrieval captures structural similarity, informalized-summary retrieval aligns semantic meaning, showing an advantage on Diffy (array manipulation tasks within \verusbench). In cases where code structures can be similar but verification targets are different, semantic differences in proof requirements make informalized summaries more effective for tasks like those in Diffy, where code index might overfit on syntax (e.g., \texttt{condm.rs} and \texttt{ms1.rs} have exactly the same sequential loop structures but require different logic reasoning—modulus operation vs. sum checking; refer to Appendix A for a detailed code comparison).

\subsection{Evaluation on \repobench}

For repository-level verification, a preliminary run of the uninformed direct generation method on the \verismo subset (first two rows in Table \ref{tab:proj_bench_result_percent}) shows its inability even under the simple tasks. We therefore focus on the performance when retrieval or refinement are incorporated. Our experiments on \repobench added the entire benchmark's corpus as the retrieval pool for few-shot examples, whereas dependency retrieval was only scoped to the tasked repository for contextual type constructs and premise lemmas. We employed code similarity for few-shot retrieval and metadata matching for dependency retrieval.


We maintained a comparable LLM budget across all experiments to ensure a fair comparison of different approaches:

\mypara{Direct Generation}
We evaluated direct proof generation (\texttt{DirectGen}) with the addition of RAG for few-shot examples and dependencies (\texttt{DirectRAG}). We generated three samples per task by default. In a greedy decoding setting, we set the LLM temperature to zero and generate only once.


\mypara{Refinement Generation}
We establish a stronger baseline by incorporating the refinement process, both with (\texttt{RAG+Refinement}) and without (\texttt{Refinement}) retrieval augmentation. These methods generated two initial samples and allowed up to two subsequent repair steps, totaling a maximum of four LLM calls.


\begin{table*}[h!]
\centering
\setlength\extrarowheight{-1pt}
\caption{Evaluation results on \repobench program verification tasks.}
\label{tab:proj_bench_result_percent}
\begin{tabular}{llcccccc}
\toprule
\textbf{Task} & \textbf{Method} & \multicolumn{2}{c}{\textbf{Correct (n, \%)}} & \multicolumn{2}{c}{\textbf{Intact (n, \%)}} & \multicolumn{2}{c}{\textbf{Success (n, \%)}} \\
& & Overall & Simple & Overall & Simple & Overall & Simple \\
\midrule
\textbf{Verismo} & & & & & & & \\
383 tasks & DirectGen greedy & N/A & 2 (3.8\%) & N/A & \textbf{52 (100.0\%)} & N/A & 2 (3.8\%) \\
52 simple tasks & DirectGen sample & N/A & 4 (7.7\%) & N/A & 48 (92.3\%) & N/A & 4 (7.7\%) \\
& Refinement & 63 (16.4\%) & 8 (15.4\%) & \textbf{294 (76.8\%)} & 49 (94.2\%) & 59 (15.4\%) & 7 (13.5\%) \\
& DirectRAG & 78 (20.4\%) & 21 (40.4\%) & 281 (73.4\%) & \textbf{52 (100.0\%)} & 65 (17.0\%) & 21 (40.4\%) \\
& RAG+Refinement & \textbf{84 (21.9\%)} & \textbf{24 (46.2\%)} & 266 (69.5\%) & 45 (86.5\%) & \textbf{75 (19.6\%)} & \textbf{23 (44.2\%)} \\
\midrule
\textbf{Ironkv} & & & & & & & \\
129 tasks & Refinement & 22 (17.1\%) & 4 (33.3\%) & 98 (76.0\%) & \textbf{11 (91.7\%)} & 21 (16.3\%) & 4 (33.3\%) \\
12 simple tasks & DirectRAG & 18 (14.0\%) & 2 (16.7\%) & \textbf{104 (80.6\%)} & 10 (83.3\%) & 18 (14.0\%) & 2 (16.7\%) \\
& RAG+refinement & \textbf{31 (24.0\%)} & \textbf{4 (33.3\%)} & 96 (74.4\%) & 10 (83.3\%) & \textbf{27 (20.9\%)} & \textbf{4 (33.3\%)} \\
\midrule
\textbf{Vest} & & & & & & & \\
\textbf{\small Vest-core} & Refinement & 17 (9.5\%) & 4 (26.7\%) & \textbf{157 (87.7\%)} & 14 (93.3\%) & 11 (6.15\%) & 4 (26.7\%) \\
179 tasks & DirectRAG & 23 (12.8\%) & 5 (33.3\%) & 152 (84.9\%) & \textbf{15 (100.0\%)} & 19 (10.6\%) & 5 (33.3\%) \\
15 simple tasks & RAG+refinement & \textbf{28 (15.6\%)} & \textbf{5 (33.3\%)} & 155 (86.6\%) & 14 (93.3\%) & \textbf{24 (13.4\%)} & \textbf{5 (33.3\%)} \\
\midrule[0.1pt]
\textbf{\small Vest-exp} & Refinement & 14 (40.0\%) & \textbf{1 (100.0\%)} & 21 (60.0\%) & \textbf{1 (100.0\%)} & 14 (40.0\%) & \textbf{1 (100.0\%)} \\
35 tasks & DirectRAG & 9 (25.7\%) & 0 (0.0\%) & 28 (80.0\%) & \textbf{1 (100.0\%)} & 9 (25.7\%) & 0 (0.0\%) \\
1 simple task & RAG+refinement & \textbf{15 (42.9\%)} & \textbf{1 (100.0\%)} & \textbf{30 (85.7\%)} & \textbf{1 (100.0\%)} & \textbf{15 (42.9\%)} & \textbf{1 (100.0\%)} \\
\midrule
\textbf{Anvil} & & & & & & & \\
29 tasks & Refinement & 9 (30.0\%) & 1 (50.0\%) & 17 (56.7\%) & \textbf{2 (100.0\%)} & 6 (20.0\%) & 1 (50.0\%) \\
2 simple tasks & DirectRAG & 10 (33.3\%) & 0 (0.0\%) & 17 (56.7\%) & \textbf{2 (100.0\%)} & 8 (26.7\%) & 0 (0.0\%) \\
& RAG+refinement & \textbf{15 (50.0\%)} & \textbf{1 (50.0\%)} & \textbf{19 (63.3\%)} & \textbf{2 (100.0\%)} & \textbf{12 (40.0\%)} & \textbf{1 (50.0\%)} \\
\bottomrule
\end{tabular}
\end{table*}

\subsubsection{Verification Results}

Repository-level verification introduces added complexity in coordinated reasoning across interdependent modules, even though our evaluation assesses proof completion per function.
We note that the Complex task setting still represents a simplification of real-world repository verification, as we focus on proving one function at a time while assuming other functions and premises are already verified. Despite this simplification, retrieval of in-distribution examples from the repository proved beneficial for successful proof completion.

The results in Table \ref{tab:proj_bench_result_percent} demonstrate a clear performance improvement across all evaluated methods when augmented with contextualized retrieval.
Yet, over 55\% of Simple tasks remained unsolved by all models, and the \repobench-Complex subset presents a more significant challenge. For instance, within the 331 Complex tasks in \repobench-\verismo, \texttt{RAG+Re-\\finement} achieved a success rate of under 16\% (52 tasks), only on par with the non-RAG \texttt{Refinement} baseline (52 tasks). This highlights the need for further advancements in both retrieval strategies and the capabilities of generative agents to tackle the complexities of repository-scale verification.

\subsubsection{Ablation Study on Retrieval Strategies}

We conducted an ablation study on the \ironkv subset of \repobench to understand the contribution of different retrieval context sources. IronKV, derived from the well-established IronFleet distributed system~\cite{hawblitzel2015ironkv}, provides a challenging yet manageable repository for this analysis. Our established baseline is the \texttt{RAG+refinement} approach, which employs a hybrid retrieval strategy: global code similarity for few-shot examples and local informalized summary similarity for dependency retrieval. The ablations are compared in Table \ref{tab:ablation_ironkv_no_simple}.

\begin{table}[ht!]
\centering
\caption{Ablation study results on Verified IronKV (Total tasks: 129).}
\setlength{\tabcolsep}{1pt}
\label{tab:ablation_ironkv_no_simple}
\centerline{
\begin{tabular}{@{}l@{\hspace{-3pt}}c@{\hspace{2pt}}c@{\hspace{0pt}}c@{\hspace{0pt}}c@{}} 
\toprule
\textbf{Ablation} & \multirow{2}{*}{\textbf{Method}} & \textbf{Correct} & \textbf{Intact} & \textbf{Success} \\
\textbf{Config.} & & \textbf{(n, \%)} & \textbf{(n, \%)} & \textbf{(n, \%)}\\
\midrule
\multirow{2}{*}{\textit{\textbf{Baselines}}} & RAG+refine & \textbf{31 (24.0\%)} & \textit{96 (74.4\%)} & \textbf{27 (20.9\%)} \\
& Refinement & \textit{22 (17.1\%)} & \textbf{98 (76.0\%)} & \textit{21 (16.3\%)} \\
\midrule
\textbf{Retrieval} & Random & \underline{20 (15.5\%)} & \textbf{100 (77.5\%)} & \underline{18 (14.0\%)} \\
\midrule
\multirow{2}{*}{\textbf{Indexing}} & Code Only & 29 (22.5\%) & 96 (74.4\%) & \textbf{28 (21.7\%)} \\
& \small{Metadata Only} & \textbf{31 (24.0\%)} & 97 (75.2\%) & \textbf{30 (23.3\%)} \\
\midrule
\textbf{RAG Scope} & Local Only & \underline{19 (14.7\%)} & \textbf{100 (77.5\%)} & \underline{16 (12.4\%)} \\
\midrule
\textbf{Dependency} & \small{GT Premise} & 23 (17.8\%) & 97 (75.2\%) & 22 (17.1\%) \\
\bottomrule
\end{tabular}
}
\end{table}

\subsubsection*{1) Random Retrieval} When all contexts were randomly retrieved, performance dropped across all metrics (\textit{Correct}, \textit{Intact}, \textit{Success}). This starkly highlights the crucial role of intelligent retrieval in providing pertinent information from the vast repository to the LLM. Randomly selected code snippets and function signatures not only fail to offer the necessary structural or semantic guidance, but also mislead the model, shown by a success rate lower than Refinement-only.

\subsubsection*{2) Index Types (Code Only vs. Metadata Only)} When only use code similarity or summary similarity for both few-shot retrieval and dependency retrieval.
    \begin{itemize}
        \item Code Only: Showed a slight decrease in \textit{Correct} rates but a marginal increase in \textit{Success}. While it leveraged structural similarity to inform correct proof practices, raw code lacked context linkages needed for dependencies.
        \item Metadata Only: Yielded slightly better performance, indicating the strong utility of semantic similarity for both examples and contextual information. Metadata summaries captured high-level intents for semantically nuanced tasks, aligning with findings from function-level benchmarks.
    \end{itemize}

\subsubsection*{3) Local-only Source} Limiting retrievals to only the tasked repository led to a substantial drop in all performance metrics. A diverse, global set of examples is essential for LLMs to learn generalizable syntax patterns and avoid overfitting on specific project conventions, which could limit their ability to leverage broader verification knowledge and successful proof strategies.

\subsubsection*{4) Ground Truth Premise} 

Solely providing ground-truth premise (function signature) lists for dependency retrieval decreased performance, indicating that LLMs require broader contextual information beyond isolated function signatures to effectively utilize dependencies. Our summary-based retrieval offered a holistic context, including usage, related type definitions, and implicit semantic relationships, which is crucial for the LLM to understand and apply the retrieved information.

In conclusion, effective repository verification with LLMs hinges on intelligent retrieval that covers a broad set of diverse examples and logical contexts. These elements collectively provide the necessary structural and semantic guidance for the LLM to navigate the complexities of repository-level reasoning.

\subsection{Qualitative Analysis of \repobench Results}
\label{appen:benchQuali}

\subsubsection{Common failure modes in \repobench}
\hspace{1pt}\newline
\repobench tasks expose two key failure modes in non-RAG-assisted models:

\begin{enumerate}
    \item \textbf{Customized Proof Dependencies}:
    Proofs often require coordinating invariants and using non-standard proof functions, which is not a convention present in common Verus tutorials.
    \item \textbf{Project-Specific Syntax Mismatches}:
    Even logically sound proofs fail due to subtle syntax deviations. Verus requires integer reasoning be performed using a ghost \texttt{Integer} type; non-RAG models frequently default to Rust’s native \texttt{i32}, causing verification failures despite correct logic.
\end{enumerate}

\subsubsection{Key difficulties faced in \repobench-Complex}
\hspace{1pt}\newline
From Table~\ref{tab:proj_bench_result_percent}, as reflected in the success rates between \texttt{\small Refine-\\ment} (59-7=52) and \texttt{RAG+Refinement} (75-23=52) on \repobench-Complex, the basic context references provided by the current retrieval methods are too simple to support much assistance in completing the hard verification tasks that require more dependencies.

The current experiments fail to handle several situations:
\begin{itemize}
    \item We observe that many contextually similar tasks in the hard category require different premise sets and proof style, requiring more specialized direction of retrieval.
    \item The ground-truth premise pool is actually larger than the maximum number of retrievals we return, demanding larger context capacity from the retrieval module.
    \item Some tasks require super long proofs (> 80 proof annotation lines). We suspect any successful run would require hundreds of refinement cycles and compiler feedbacks, in addition to a fully complete premise pool, which is out of our current sampling budget.
\end{itemize}

\subsubsection{Generation Style}
\hspace{1pt}\newline
Since the number of verification success, as a binary metric, does not capture gradual improvement in the proof generation quality using LLMs, we evaluate the code similarity with respect to the ground truth proof to reflect how much of the proofs are in the right directions. We analyze the average BLEU scores between the generated code and the ground-truth code for selected pipeline settings.

\begin{table}[h!]
\centering
\caption{BLEU scores between generated answers and the ground-truth function}
\label{table:bleu_scores}
\begin{tabular}{>{\bfseries}l c}
\toprule
\textbf{Proof Source} & \textbf{Average BLEU Score ($\uparrow$)} \\
\midrule
Code without Proof & 46.76 \\
\midrule
DirectGen Baseline & 46.32 \\
Refinement Baseline & 48.18 \\ 
\midrule
DirectRAG & \textbf{57.75} \\
RAG+Refinement & 55.97 \\
\bottomrule
\end{tabular}
\end{table}

We observe that both retrieval augmented pipelines produce more coherent answers, suggesting that they are utilizing the project-specific contexts as expected.

\section{Related Work}

\mylist{Automated program verification with LLM} While there exist various techniques for automated program verification with machine learning based methods like Code2Inv~\cite{si2020code2inv}, CIDER~\cite{liu2022learning} and Code2RelInv~\cite{wang2022learning}, there has been recent advancements focusing on the integration of LLMs to enhance proof generation capabilities~\cite{wen2024enchanting}. LLMs provide the potential to automate these processes by generating human-like proofs~\cite{wen2024enchanting} and code~\cite{li2024guiding}. LLM-aided proof/code generation has been studied within verification-aware programming languages like Frama-C~\cite{kirchner2015frama}, Dafny~\cite{leino2010dafny} and Verus~\cite{lattuada2023verus}. Recent work \autoverus ~\cite{yang2024autoverus} leverages LLMs for automated proof synthesis in Rust using finetuned knowledge bases and refinement processes. Clover~\cite{sun2024clover} addresses the case of software verification where no specification is formally given, and attempts to autoformalize a specification in Dafny by aligning with the available implementations and documentations. LeanDojo~\cite{yang2023leandojo} introduces a large premise pool in Lean and uses fine-tuned retrieval models to perform RAG for proofs. Although these approaches differ in methodology and application areas, they all focus primarily on single-function verification. 

\mypara{Repository-level program verification} Repository-level program verification is a relatively emerging field with limited previous research addressing the inherent complexity of large software systems. Selene~\cite{zhang2024selene}, tailored on verifying the seL4 microkernel using the Isabelle theorem proving language, represents a pioneering effort in this domain, but the language focus, verification tooling, and dependency management approaches are different from \ragverus. While there has been recent development in repository-level LLM-based code generation~\cite{liao20243}, \ragverus extends this domain into automated verification, ensuring formal correctness.

\section{Future Work}

Building upon the \ragverus framework and the \repobench benchmark, future research can focus on enhancing the scalability, precision, and intrinsic capabilities of LLM-driven verification for repository-level tasks. Key directions include:

\mylist{Scaling Verification Beyond Individual Completions} Current approaches assume dependencies are pre-filled and pre-verified. Future work should extend verification to handle concurrent proof synthesis across multiple interdependent modules simultaneously. Completing specification, implementation, and verification in a single stack bridges the gap between benchmarks and the real-world codebases.

\mylist{Specializing Retrieval for Structural and Semantic Dependencies} While \ragverus demonstrates the value of retrieval, significant challenges remain in accurately identifying complex premises and project-specific contexts. More fine-grained retrieval methods could involve finetuning embedding encoders to capture causal relations between callers and callees in code dependencies, moving beyond generic similarity metrics to retrieve essential type definitions, lemmas, and patterns for complex proof construction.

\mylist{Improving Intrinsic LLM Capabilities for Verification Languages} Verification languages like Verus present unique syntactic and logical challenges for general-purpose LLMs. A promising direction is to enhance the intrinsic capabilities of LLMs in these niche domains. This can be achieved through fine-tuning on verification-specific corpora and leveraging feedback mechanisms from the compiler/verifier for reinforcement learning, enabling models to learn the verification syntax and improving their reasoning on the topic.


\section{Conclusion}

This paper tackles the challenge of scaling formal verification to repository-level by introducing \repobench, the first benchmark specifically designed for repository verification tasks in Verus, and \ragverus, a framework leveraging context-aware prompting to enhance LLM-based proof synthesis.
By focusing on repository-level challenges, our work establishes foundational steps towards practical, automated verification suitable for large-scale, real-world software systems.

\section*{Acknowledgements}

This work was supported, in part, by Individual Discovery Grant program from the NSERC of Canada and the Canada CIFAR AI Chair Program. This work has also benefitted from the Microsoft Accelerate Foundation Models Research (AFMR) grant program.
We thank Jiading Zhu and Yifang Tian for their early exploration on the idea, and Zhiyang Chen for his valuable feedback and revision suggestions.

\appendix
\section*{Appendix}

\section{Code Simiarity vs. Semantic Similarity}

In \verusbench-Diffy, different problems are hard to differentiate by their code syntax. However, we still need to target individual problems to provide suitable guidance for proof completion. As an example, two functions \texttt{condm.rs} and \texttt{ms1.rs} both have the following serial-loop structure,

\input{code/fmt} 
{\scriptsize
\begin{Verbatim}[commandchars=\\\{\},codes={\catcode`\$=3\catcode`\^=7\catcode`\_=8\relax}]
\PY{+w}{  }\PY{l+m+mi}{1}\PY{+w}{ }\PY{k+kd}{let}\PY{+w}{ }\PY{k}{mut}\PY{+w}{ }\PY{n}{i}\PY{p}{:}\PY{+w}{ }\PY{k+kt}{usize}\PY{+w}{ }\PY{o}{=}\PY{+w}{ }\PY{l+m+mi}{0}\PY{p}{;}
\PY{+w}{  }\PY{l+m+mi}{2}\PY{+w}{ }\PY{k}{while}\PY{+w}{ }\PY{p}{(}\PY{n}{i}\PY{+w}{ }\PY{o}{\PYZlt{}}\PY{+w}{ }\PY{n}{N}\PY{+w}{ }\PY{k}{as}\PY{+w}{ }\PY{k+kt}{usize}\PY{p}{)}\PY{+w}{ }\PY{p}{\PYZob{}}
\PY{+w}{  }\PY{l+m+mi}{3}\PY{+w}{   }\PY{o}{..}\PY{p}{.}
\PY{+w}{  }\PY{l+m+mi}{4}\PY{+w}{   }\PY{n}{i}\PY{+w}{ }\PY{o}{=}\PY{+w}{ }\PY{n}{i}\PY{+w}{ }\PY{o}{+}\PY{+w}{ }\PY{l+m+mi}{1}\PY{p}{;}
\PY{+w}{  }\PY{l+m+mi}{5}\PY{+w}{ }\PY{p}{\PYZcb{}}
\PY{+w}{  }\PY{l+m+mi}{6}\PY{+w}{ }\PY{n}{i}\PY{+w}{ }\PY{o}{=}\PY{+w}{ }\PY{l+m+mi}{0}\PY{p}{;}
\PY{+w}{  }\PY{l+m+mi}{7}\PY{+w}{ }\PY{k}{while}\PY{+w}{ }\PY{p}{(}\PY{n}{i}\PY{+w}{ }\PY{o}{\PYZlt{}}\PY{+w}{ }\PY{n}{N}\PY{+w}{ }\PY{k}{as}\PY{+w}{ }\PY{k+kt}{usize}\PY{p}{)}\PY{+w}{ }\PY{p}{\PYZob{}}
\PY{+w}{  }\PY{l+m+mi}{8}\PY{+w}{   }\PY{k}{if}\PY{p}{(}\PY{o}{..}\PY{p}{.}\PY{p}{)}\PY{+w}{ }\PY{p}{\PYZob{}}\PY{o}{..}\PY{p}{.}\PY{p}{\PYZcb{}}
\PY{+w}{  }\PY{l+m+mi}{9}\PY{+w}{   }\PY{k}{else}\PY{+w}{ }\PY{p}{\PYZob{}}\PY{+w}{ }\PY{o}{..}\PY{p}{.}\PY{+w}{ }\PY{p}{\PYZcb{}}
\PY{+w}{ }\PY{l+m+mi}{10}\PY{+w}{   }\PY{n}{i}\PY{+w}{ }\PY{o}{=}\PY{+w}{ }\PY{n}{i}\PY{+w}{ }\PY{o}{+}\PY{+w}{ }\PY{l+m+mi}{1}\PY{p}{;}
\PY{+w}{ }\PY{l+m+mi}{11}\PY{+w}{ }\PY{p}{\PYZcb{}}
\end{Verbatim}

}



\noindent
making their code look similar at the syntax level.
However, their required proofs and invariants are actually different; one deals with element selection, and the other handles numeric summation, thus requiring different reasoning guidance.

Metadata summaries recognize the functional differences and can be used to retrieve related examples effectively. As shown in Figure~\ref{fig:metadata_comparison}, the summaries explicitly identify that \texttt{condm.rs} deals with modulus operation at individual indices, while \texttt{ms1.rs} focuses on summation over the list.

\begin{figure*}[htbp]
    \centering
    \hspace{0.03\textwidth} 
    \begin{minipage}[t]{0.45\textwidth}
        \lstset{language=Python, basicstyle=\ttfamily\scriptsize, backgroundcolor={}}
        \begin{lstlisting}
...
### Functionality Summary:
1. **Loop 1**: Iterates over the vector from index `0` to `N-1`, setting each element to `0`.
2. **Loop 2**: Iterates over the vector again from index `0` to `N-1`. If `N` is even, it increments each element by `2`; if `N` is odd, it increments each element by `1`.
3. **Post-condition**: After the second loop, each element in `a` will have a value such that `a[k] % 2` is equal to `N % 2`.

...

### Invoked Functions:

- `a.set(i, value)`: This function sets the element at index `i` of the vector `a` to `value`. This is a method of the `Vec` type in Rust or a custom method provided by the `verus` library.


        \end{lstlisting}
    \end{minipage}
    \hspace{0.05\textwidth} 
    \begin{minipage}[t]{0.45\textwidth}
        \lstset{language=Python, basicstyle=\ttfamily\scriptsize, backgroundcolor={}}
        \begin{lstlisting}
...
### Functionality Summary:
  - If `i` is 0, it sets the first element of `sum` to 0.
  - For all other indices, it adds the current element of `a` to the first element of `sum`. However, since all elements of `a` are 0, `sum[0]` remains 0.
  - The function ensures that the first element of `sum` is 0 at the end of execution, which is consistent with the specified `ensures` condition.
...

### Invoked Functions:
- `a.set(i, i % 1)`: This is a method call on the vector `a` to set the element at index `i` to `i % 1`.
- `sum.set(0, 0)`: This is a method call on the vector `sum` to set the element at index 0 to 0.
- `sum.set(0, sum[0] + a[i])`: This is a method call on the vector `sum` to update the element at index 0 by adding the current element of `a` at index `i`.
        \end{lstlisting}
    \end{minipage}
    \vspace{-\baselineskip}
    \caption{Comparison of \textbf{metadata summaries} for \texttt{condm.rs} (left) and \texttt{ms1.rs} (right), where the informalized summaries explicitly identify the logic differences and the key objects being operated on.}
    \label{fig:metadata_comparison}
\end{figure*}

\section{Data Processing during \ragverus}
\label{appendix:data}

For any tasked repository, we formulate all available contents for retrieval before accepting a problem. We preprocess 1) the code for all functions, 2) metadata linking types and constructs, and 3) informalized semantic summaries.

\mypara{Mining Code Properties}
We process repository-wide code artifacts to extract verification-critical metadata following Section 3.2, building on insights from context modeling in repository-level code generation~\cite{shrivastava2023repoPrompt}.
\begin{itemize}
    \item Context types: We identify function/type signatures, method calls, type identifiers and field declarations with in the target function
    \item Type sources: We look for the source definitions of the appearing context types from the current file, parent/child classes, and imports.
\end{itemize}
For each function, we collect these contexts across the repository to build a linkage graph by analyzing the control/dataflow via static analysis of the codebase hierarchy, reflecting module dependencies.

\mypara{Informalizing Code Summaries}
This stage generates natural language descriptions of Rust functions, termed informalized semantic summaries, to supplement the contexts. Using the code implementation and the processed metadata, we produce concise summaries using GPT-4o to encapsulate the function’s purpose, behavior, and key properties. For example, a function performing cryptographic hashing can be summarized as: "\textit{Computes a SHA-256 hash of input bytes, enforcing non-null pointers and initializing a secure context}..."
Informalization-based example selection leverages these summaries during retrieval, matching them by semantic similarity to prioritize functionally analogous examples. This ensures retrieved contexts align with the target’s verification intent rather than superficial syntax, addressing where syntax and static analysis fails.

\begin{figure}[H]
    \setlength{\abovecaptionskip}{0pt}
    \setlength{\belowcaptionskip}{-10pt}
    \centering
    \includegraphics[width=0.48\textwidth]{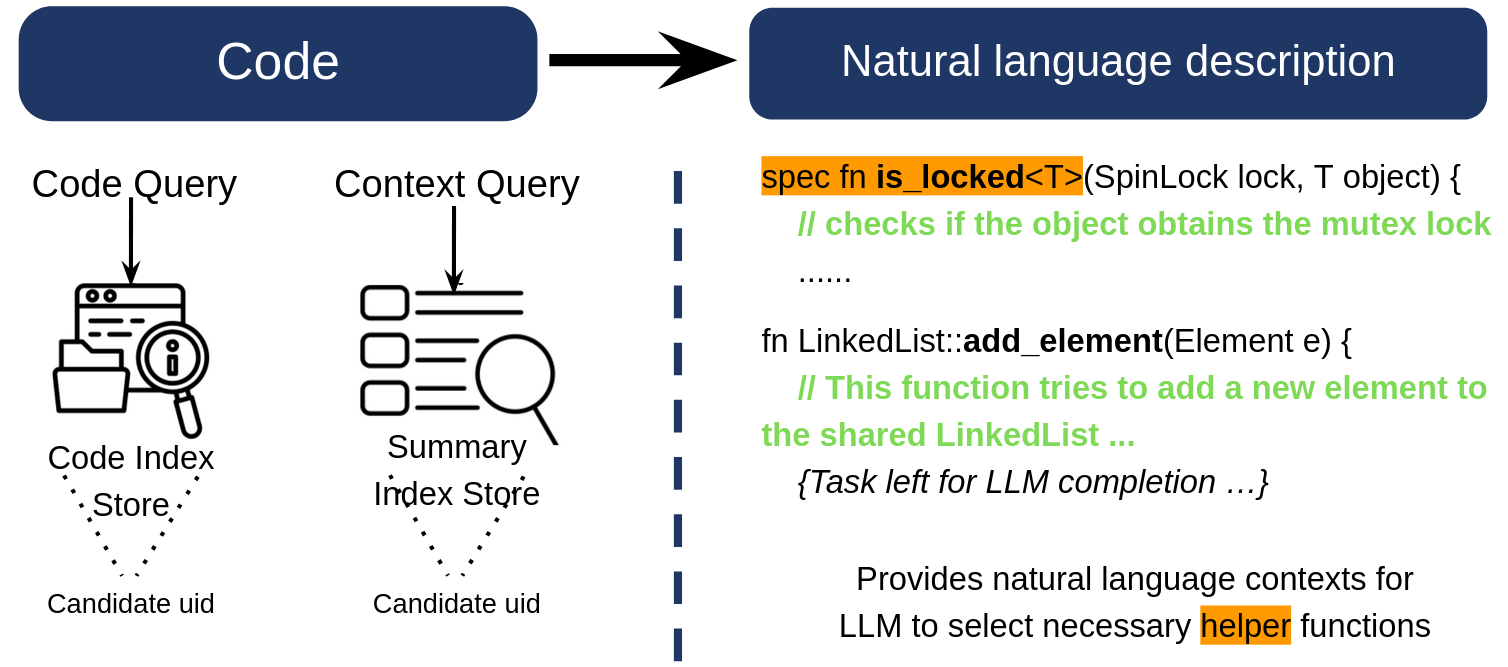}
    \caption{Code informalization summarizes the code functionalities. Here, one can identify through the {\color{LimeGreen}summaries} that \texttt{is\_locked} is a good helper candidate for the task \texttt{add\_element}.} 
    \label{fig:code_informalization}
\end{figure}

\mypara{Generating Vectorized Indices}
To enable efficient retrieval, \ragverus uses persistent vector stores where documents are encoded into embeddings using OpenAI's \texttt{text-\\embedding-3-large} model~\cite{openai2024embedding}.
We encode the Rust code and their informalized summaries separately into high-dimen-sional embeddings; these embeddings are then indexed via FAISS~\cite{faisspaper} (implemented through LlamaIndex~\cite{llamaindex}), a library optimized for fast similarity search across large corpora. Two distinct indices are created for each kind of context sources:
\begin{itemize}[noitemsep,topsep=0pt,parsep=0pt,partopsep=0pt]
    \item \textbf{Unverified Index}: Contains embeddings of unverified code and summaries, aligning with the input queries during proof synthesis.
    \item \textbf{Verified Index}: Stores fully verified code and their complete semantics, preserving ground-truth context for validation.
\end{itemize}
These indices capture both syntactic and semantic patterns and are stored persistently, enabling reuse across experiments without recomputing embeddings.


\mypara{Handling Queries}
Relevant contexts of a targeted task can be retrieved by querying the FAISS indices with the task's code or summary, where the item with the smallest embedding distance to the target is returned.
To prevent data leakage during experiment, we filter the task's own context if found during retrieval, ensuring proofs are synthesized solely from external dependencies for the unverified task.

\mypara{Addressing Repository Context in Verification}
\textsc{Rag}-\textsc{Verus} effectively integrates the aforementioned semantic embeddings and static code analysis to prioritize domain-specific syntax (e.g., type definitions) and repository structure (e.g., dependencies). This ensures precise retrieval of relevant lemmas and invariants, avoiding over-retrieval while grounding LLM-generated proofs in verified project contexts.

\bibliographystyle{ACM-Reference-Format}

\end{document}